\documentclass[a4paper,11pt]{article}
\usepackage{pos}

\title{The ALICE asynchronous software trigger processing for LHC Run 3}
\ShortTitle{ALICE asynchronous software triggers in LHC Run 3}

\author*[a]{Sara Pucillo}
\onbehalf{for the ALICE Collaboration}

\affiliation[a]{Università degli Studi di Salerno and Sezione INFN Salerno,\\
  Via Giovanni Paolo II, 132, Fisciano (SA), 84084, Italy}
\emailAdd{sara.pucillo@cern.ch}

\abstract{
The detector and computing upgrades for LHC Run~3 enable ALICE to continuously acquire proton--proton (pp) and lead--lead (Pb--Pb) collisions at interaction rates of up to 1.5~MHz and 50~kHz, respectively. The resulting pp dataset is more than three orders of magnitude larger than the combined Run~1 and Run~2 samples, and its long-term retention in full would be impractical given the available storage capacity.

To address this challenge, ALICE has implemented a physics-driven asynchronous software trigger designed to retain only collisions containing signals of physics interest. Following full asynchronous reconstruction, the Central Event Filter Processor applies more than 100 analysis-based selection channels spanning the ALICE pp physics programme. The selected collisions are then used to skim the original Compressed Time Frames (CTFs), preserving the relevant detector information within detector-specific time windows. This procedure reduces the retained CTF volume to approximately 4.5\% of its original size, while preserving the possibility of future reconstruction with improved calibrations and software.

This contribution describes the implementation and operation of the asynchronous software trigger, from collision selection to CTF skimming, and highlights how this strategy provides access to rare signals and processes across the ALICE physics programme.
}

\FullConference{43rd International Conference on High Energy Physics (ICHEP 2026)\\
30 July to 5 August, 2026\\
Natal, Brazil}

\hypersetup{
  pdftitle={The ALICE asynchronous software trigger processing for LHC Run 3},
  pdfauthor={Sara Pucillo for the ALICE Collaboration},
  pdfsubject={ICHEP 2026 proceedings},
  pdfkeywords={ALICE, Run 3, asynchronous software trigger, continuous readout, CTF skimming}
}

\begin{document}
\maketitle

\section{ALICE in Run 3}
The ALICE Run~3 physics programme aims at high-precision measurements of the quark--gluon plasma, with particular emphasis on rare and low-transverse-momentum ($p_{\rm T}$) probes. Studies of heavy-flavour hadrons, quarkonia, strange hadrons, light nuclei and hypernuclei, as well as femtoscopic correlations, require large collision samples and improved tracking and vertexing performance. To meet these requirements, ALICE underwent major detector upgrades and transitioned to continuous data acquisition~\cite{ALICEUpgrade}. 

The upgraded Inner Tracking System (ITS2) consists of seven layers of silicon-pixel detectors, with the innermost layer moved closer to the interaction point, from approximately 4~cm to 2.2~cm. This reduced distance, together with other detector improvements, enhances the pointing resolution by up to a factor of two in the transverse plane and five along the beam direction. In the Time Projection Chamber (TPC), gas-electron-multiplier (GEM) readout chambers replaced the previous multi-wire proportional chambers, enabling continuous operation. The Muon Forward Tracker provides vertexing capabilities for forward muons, while the Fast Interaction Trigger supplies precise timing, luminosity and multiplicity information. Together with upgraded readout electronics, these detectors enable continuous data acquisition at interaction rates of approximately 50~kHz in Pb--Pb collisions and up to 1.5~MHz in pp collisions.

\begin{figure}[!htbp]
  \centering
  \includegraphics[width=0.49\textwidth]{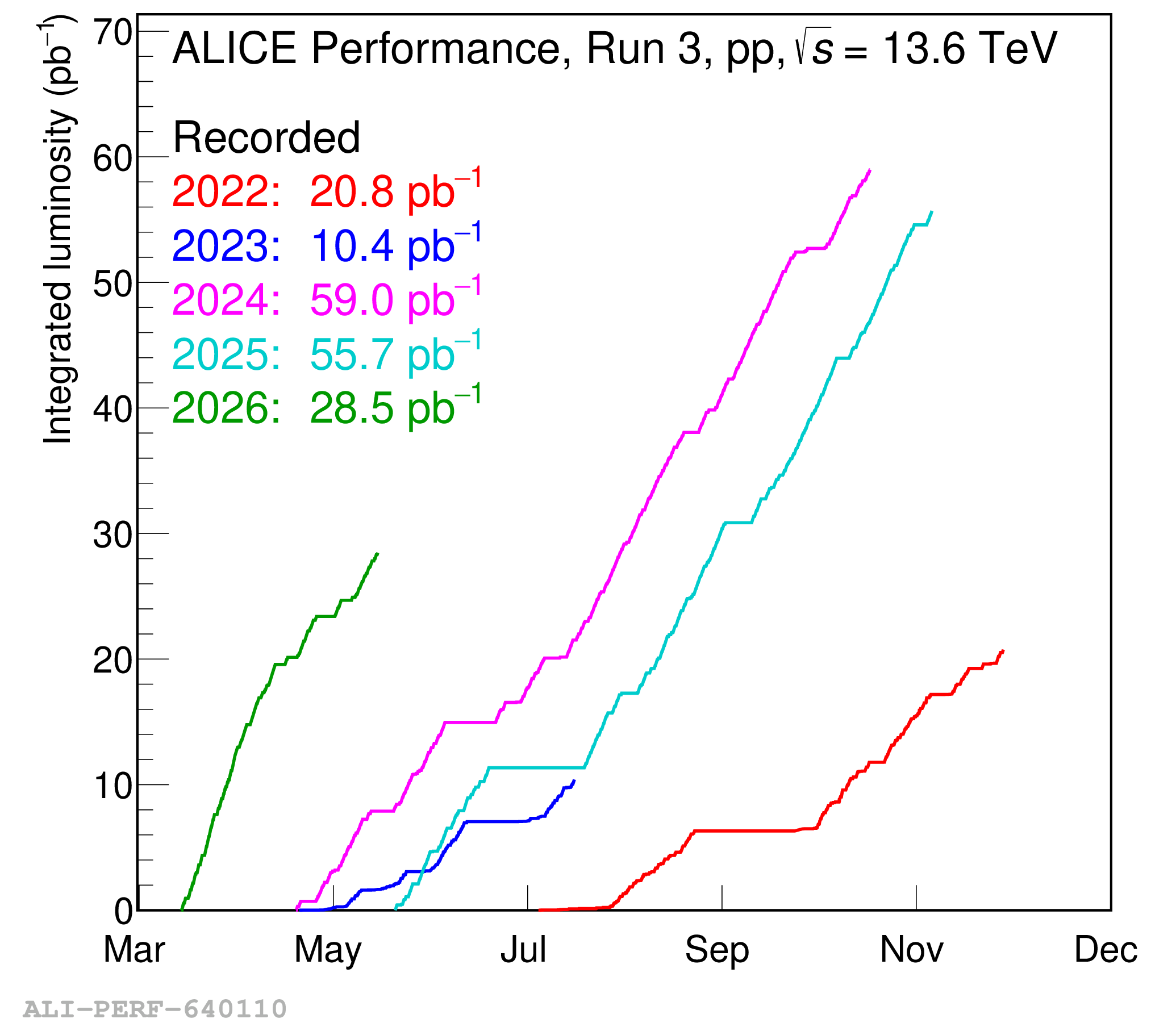}
  \includegraphics[width=0.49\textwidth]{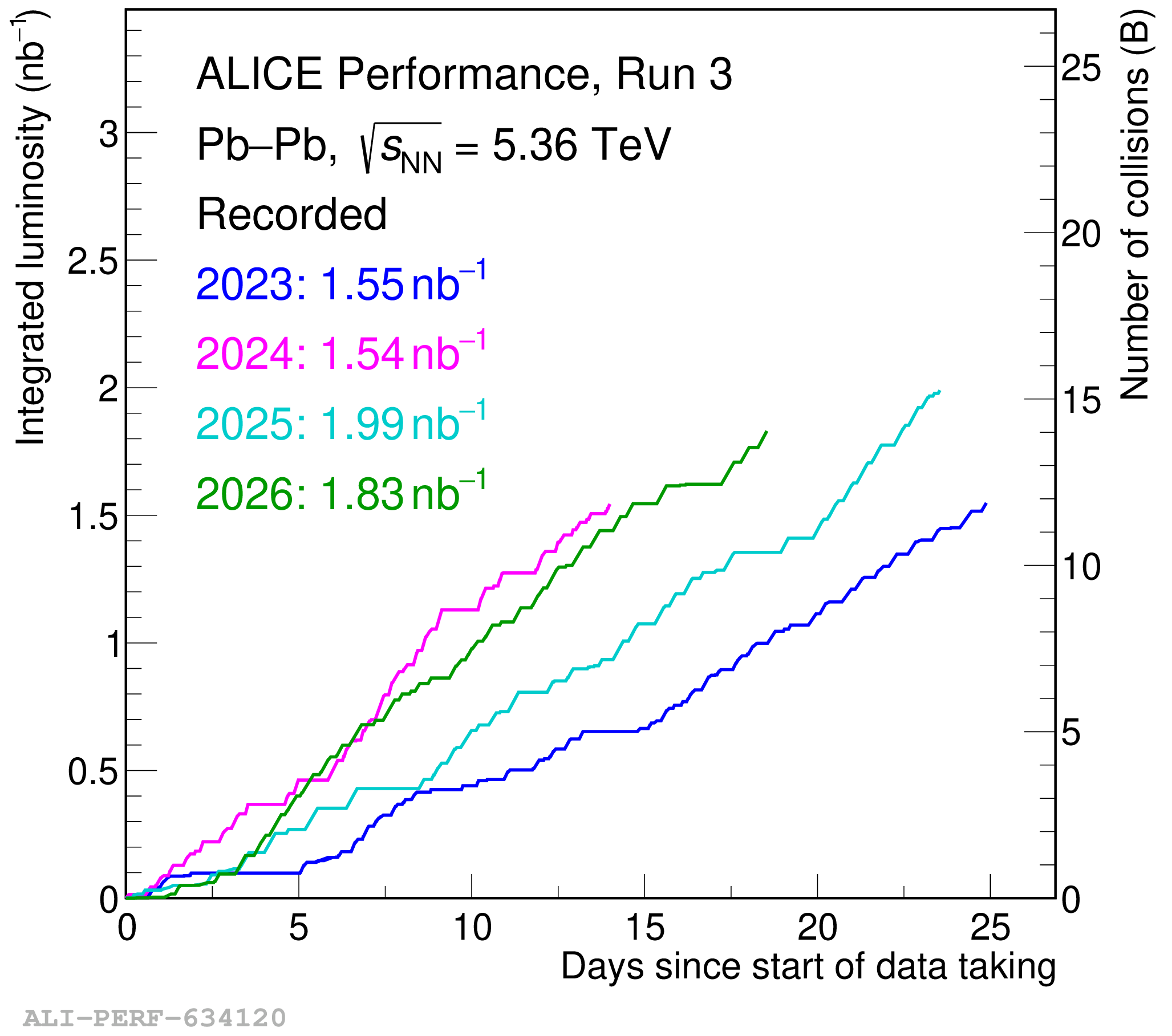}
  \caption{Cumulative integrated luminosity recorded by ALICE in pp collisions at $\sqrt{s}=13.6$~TeV (\textit{left}) and Pb--Pb collisions at $\sqrt{s_{\rm NN}}=5.36$~TeV (\textit{right}), shown separately for each year of Run~3.}
  \label{fig:luminosity}
\end{figure}

Figure~\ref{fig:luminosity} shows the integrated luminosity recorded by ALICE in pp collisions at $\sqrt{s}=13.6$~TeV and Pb--Pb collisions at $\sqrt{s_{\rm NN}}=5.36$~TeV. During Run~3, ALICE collected more than 170~pb$^{-1}$ of integrated luminosity in pp collisions, corresponding to approximately $9.5\times10^{12}$ minimum-bias events, more than three orders of magnitude larger than the combined Run~1 and Run~2 samples. However, storing the full Run~3 pp dataset would require approximately 600~PB, exceeding the available long-term storage capacity. To overcome this limitation, ALICE introduced a physics-driven asynchronous software trigger strategy, which selects collisions containing signals of physics interest and retains only the corresponding detector information, significantly reducing the volume of data to be preserved.

\section{Continuous readout and raw-data flow}
The integrated online--offline ($\mathrm{O}^{2}$) computing system organises continuous detector data into time frames\footnote{A time frame is a fixed-duration interval of continuous detector readout, containing data from multiple collisions.}~\cite{O2}. Approximately 200 First Level Processors (FLPs), directly connected to the detectors, collect, aggregate and preprocess the incoming data stream. The data are then transferred to approximately 350 GPU-equipped Event Processing Nodes (EPNs), which perform synchronous tracking, vertexing, calibration and compression~\cite{GPU}. A chain of processing and compression stages distributed across the two computing farms progressively reduces the data rate during acquisition. For 50~kHz Pb--Pb operation, the initial detector stream of approximately 3.5~TB/s is reduced to about 700~GB/s before being transferred from the FLPs to the EPNs. Further reconstruction and compression at the EPNs reduce the output to approximately 130~GB/s, corresponding to an overall data-rate reduction by more than a factor of 20. In pp running, the output to disk is approximately 50~GB/s.

The resulting Compressed Time Frames (CTFs) are written to a disk buffer with a capacity of approximately 150~PB, where they are temporarily stored for subsequent asynchronous processing. This buffering separates real-time acquisition from asynchronous reconstruction, allowing the latter to be performed after data taking, using final calibration parameters and detector conditions.

Despite the substantial reduction achieved during synchronous processing, the full pp data stream cannot be retained within the available long-term storage capacity. The estimated storage requirements for the complete Run~3 datasets amount to approximately 186~PB for Pb--Pb and 600~PB for pp collisions. While the entire Pb--Pb dataset can be preserved, the much larger pp data volume requires an additional physics-driven reduction. For this purpose, ALICE introduced an asynchronous software trigger that identifies collisions of physics interest using fully reconstructed data. The corresponding detector information is then retained through a skimming procedure applied to the original CTFs, as described in the following sections.

\section{Asynchronous reconstruction and CTF skimming}

The first stage of asynchronous reconstruction is common to pp and Pb--Pb collisions. After data taking, the CTFs stored in the disk buffer are processed using EPN and distributed Grid resources. Unlike synchronous processing, this stage is not constrained by the real-time acquisition rate and can therefore use final calibration parameters and detector conditions to perform the full reconstruction of tracks and collisions, producing analysis-ready Analysis Object Data (AO2D) tables.

For pp collisions, an additional stage is introduced. The Central Event Filter Processor (CEFP) runs a set of analysis tasks that identify collisions containing potentially interesting physics signals, according to the predefined trigger menu discussed in Section~\ref{sec:menu}. These offline tasks implement selections targeting specific signals or topologies, such as heavy-flavour hadrons, jets, nuclei and multi-strange particles. Their output is a decision table containing one column for each trigger channel, indicating whether a given collision satisfies the corresponding selection. The selected collisions are then associated with their corresponding bunch crossings (BCs), providing the time references needed for the subsequent skimming procedure.

Starting from this event-tagging information, the original CTFs are skimmed to retain only the detector data relevant to the selected collisions. However, this operation is more complex than simply keeping the data recorded at the bunch crossing of interest. Different detectors have different time responses, readout structures and timing offsets, while the reconstructed collision time has a finite resolution. Consequently, a dedicated BC window is defined around each tagged BC for every detector, taking these characteristics into account.

For example, the TPC records signals in time bins of approximately 200~ns, and the selected data must cover the full electron drift time (100 $\mu$s). The Inner Tracking System (ITS) and Transition Radiation Detector (TRD) instead organise their data into readout frames of approximately 5~$\mu$s and 3~$\mu$s, respectively. Hits recorded by the TOF detector are time-ordered and must be corrected for the readout time offset before use. The detector-specific selection windows therefore differ in extent and position, with an additional tolerance of order 1000~BCs to account for timing uncertainties and offsets. All detector information falling outside the selected ranges is removed from the original CTFs. In this way, the skimming procedure preserves the information needed to reconstruct the selected collisions while substantially reducing the data volume.

The resulting skimmed CTFs replace the original ones in long-term storage, following the necessary quality-assurance checks. A further asynchronous reconstruction pass is performed on the skimmed data to produce the final AO2Ds used for physics analysis. Importantly, the archived CTFs remain fully reconstructable, allowing future reconstruction passes to exploit improved calibrations, detector conditions and software. Thus, the data volume is significantly reduced without losing the possibility of reprocessing the selected collisions.

The complete processing workflow, including calibration, asynchronous reconstruction, offline trigger selection, CTF skimming and quality assurance, is organised to process a batch corresponding to approximately one week of data taking within four weeks. This allows analysis-ready datasets to become available within a few weeks of data acquisition.

\section{Trigger menu and access to rare signals}\label{sec:menu}
\begin{figure}[!htbp]
  \centering
  \includegraphics[angle=90,width=0.75\textheight]{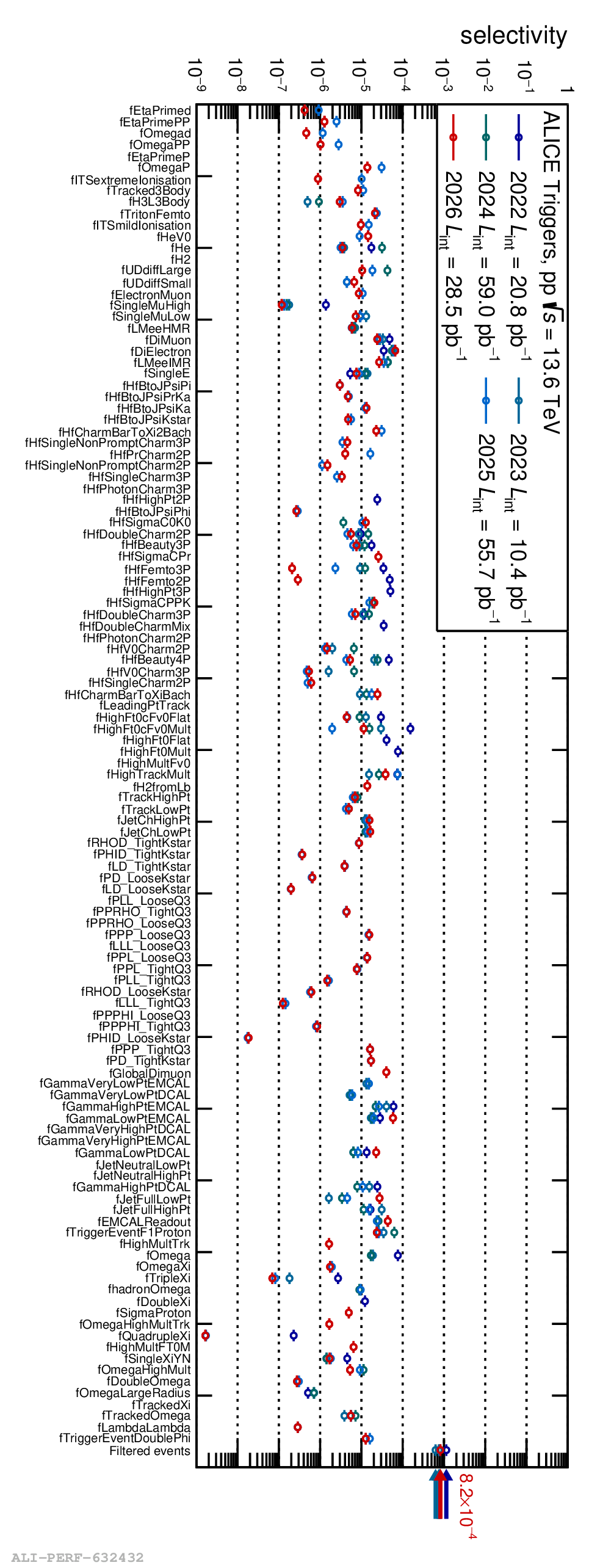}
  \caption{Selectivity of the ALICE software-trigger channels in pp collisions at $\sqrt{s}=13.6$~TeV for the 2022--2026 samples, and the combined selection at the right. The integrated luminosities in the legend refer to the samples used for this trigger-performance plot. The rightmost point, marked by the arrow, shows the total fraction of collisions selected by the combination of all trigger channels.}
  \label{fig:menu}
\end{figure}

The Run~3 menu contains more than 100 channels covering dileptons and quarkonia, light and heavy flavour, correlations and flow, jets, electromagnetic probes, and ultraperipheral and diffractive processes. Figure~\ref{fig:menu} shows their selectivities for the 2022--2026 pp samples. The selectivity of a channel is the number of collisions tagged by that channel divided by the total number of analysed collisions. The values span several orders of magnitude, reflecting the different rates of the targeted signatures. Some channels are additionally downscaled to control the number of selected events. The total selected-collision fraction is below $10^{-3}$. After applying the complete trigger selection and skimming procedure, approximately 4.5\% of the original CTF volume is retained.

The software-trigger selection, followed by subsequent reconstruction and analysis of the skimmed data, provides access to rare signals and processes across the ALICE physics programme, as illustrated by the examples discussed below.

\begin{figure}[!htbp]
  \centering
  \includegraphics[height=5.9cm,keepaspectratio]{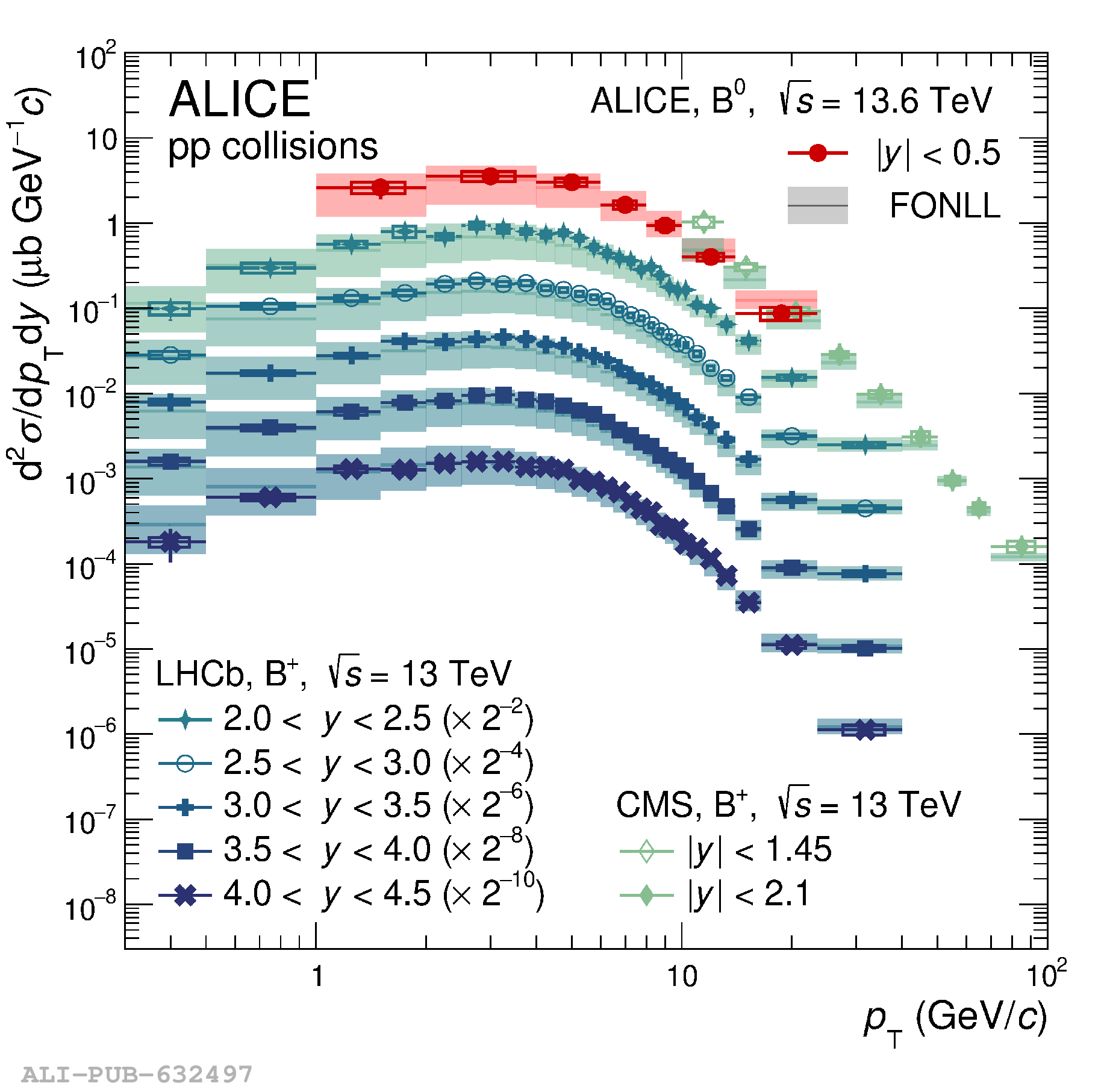}
  \includegraphics[height=5.9cm,keepaspectratio]{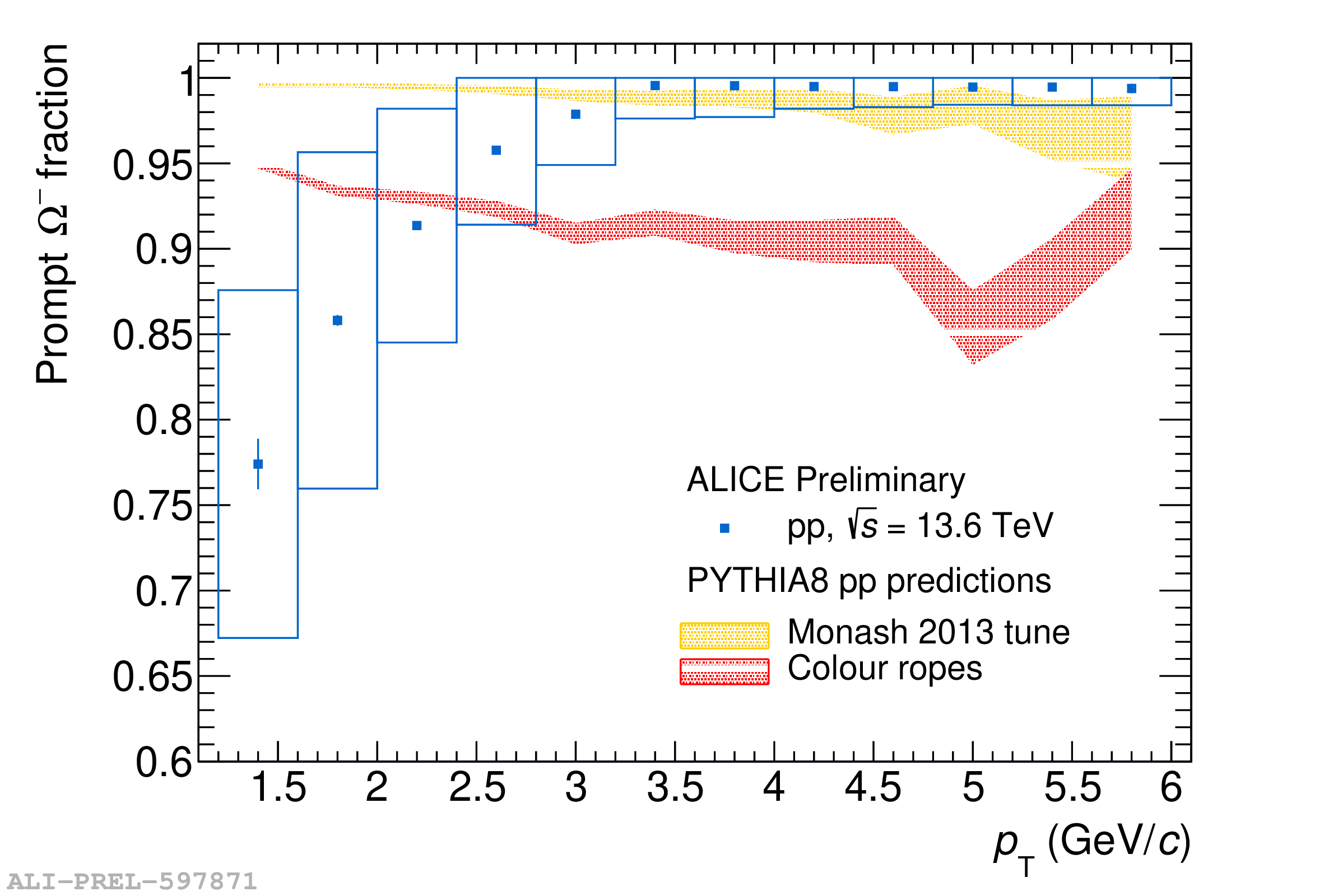}
  \caption{Examples of measurements enabled by the Run~3 pp software-trigger strategy. \textit{Left}: $\mathrm{B}^{0}$-meson production cross section at midrapidity compared with $\mathrm{B}^{+}$ measurements at other rapidities by CMS and LHCb~\cite{B0}. \textit{Right}: preliminary prompt $\Omega^{-}$ fraction as a function of $p_{\rm T}$, compared with PYTHIA~8 predictions using the Monash and colour-rope configurations~\cite{promptomega}.}
  \label{fig:physicsresults}
\end{figure}

Dedicated beauty-hadron selections enabled the first measurement of the $\mathrm{B}^{0}$-meson production cross section at midrapidity ($|y|<0.5$) down to $p_{\rm T}=1~\mathrm{GeV}/c$ at LHC energies~\cite{B0}. The mesons were reconstructed via their $\mathrm{B}^{0}\to\mathrm{D}^{-}\pi^{+}$ decay channel, followed by $\mathrm{D}^{-}\to\mathrm{K}^{+}\pi^{-}\pi^{-}$. Figure~\ref{fig:physicsresults} (left) illustrates the extended low-$p_{\rm T}$ reach of ALICE with respect to earlier beauty-hadron measurements by CMS, performed in different rapidity regions. The measured cross section is described within uncertainties by perturbative-QCD calculations. A complementary $\mathrm{B}^{+}\to J/\psi\,\mathrm{K}^{+}$ analysis also probes beauty production at low $p_{\rm T}$.

Another example combines physics-driven selections with the ITS2 strangeness-tracking algorithm. By matching cascade topologies reconstructed in the outer ITS2 layers to tracks in the innermost layers, the algorithm improves the reconstruction of strange baryons and the resolution of their distance of closest approach (DCA) to the primary vertex. The prompt $\Omega^{-}$ fraction is extracted from a simultaneous fit to the transverse and longitudinal DCA distributions, including prompt, charm-feed-down and beauty-feed-down components~\cite{promptomega}. The preliminary result, shown in Fig.~\ref{fig:physicsresults} (right), suggests a non-prompt contribution of order 20\% at low $p_{\rm T}$, although uncertainties associated with poorly known charm-baryon decay modes remain substantial. The measurement provides a way to investigate the role of heavy-flavour decays in strange-baryon production and, in the future, its multiplicity dependence.

Dedicated nuclear channels provide access to even rarer composite states. The anti-hypertriton sample offers approximately 40 times the Run~2 statistics used for the earlier measurement~\cite{hypertriton}, enabling a more precise $p_{\rm T}$ spectrum and tests of coalescence models and the anti-hypertriton wave function. The first observation of anti-$^{4}$He in pp collisions was also feasible. Correlation triggers extend the reach to three-body femtoscopy, with up to 40 times more triplets than in Run~2~\cite{ppLambda,ppK}, enabling for the first time access to the pp$\Xi$ system. These examples illustrate how the same selection and skimming chain supports physics channels spanning very different production rates.

\section{Summary and outlook}
Continuous acquisition allows ALICE to inspect a pp collision sample far larger than those available in the previous LHC runs. The asynchronous software trigger turns this increase into usable physics data by selecting collisions after full reconstruction and preserving the detector-specific time intervals required to reconstruct them again. More than 100 channels retain rare signatures while reducing the stored pp CTF volume to approximately 4.5\% of its original size.

The resulting samples support measurements of beauty and charm production, strange hadrons, rare (hyper)nuclei and multi-particle correlations. Their availability within weeks of data taking, together with the possibility of reprocessing the skimmed CTFs, links prompt analysis to long-term data preservation. For Run~4, larger datasets and further automation of the trigger and validation workflow are expected to extend this programme.


\begin{thebibliography}{9}
\setlength{\itemsep}{0pt}
\setlength{\parsep}{0pt}
\setlength{\topsep}{0pt}
\small
\bibitem{ALICEUpgrade}
ALICE Collaboration, \emph{ALICE upgrades during the LHC Long Shutdown 2}, JINST \textbf{19} (2024) P05062, \href{https://doi.org/10.1088/1748-0221/19/05/P05062}{doi:10.1088/1748-0221/19/05/P05062}.

\bibitem{O2}
ALICE Collaboration, \emph{Technical Design Report for the Upgrade of the Online--Offline Computing System}, CERN-LHCC-2015-006, ALICE-TDR-019 (2015).

\bibitem{GPU}
F.~Ronchetti \emph{et al.}, \emph{Efficient high performance computing with the ALICE event processing nodes GPU-based farm}, Front. Phys. \textbf{13} (2025) 1541854, \href{https://doi.org/10.3389/fphy.2025.1541854}{doi:10.3389/fphy.2025.1541854}.

\bibitem{B0}
ALICE Collaboration, \emph{Measurement of the $\mathrm{B}^{0}$-meson production cross section in proton--proton collisions at $\sqrt{s}=13.6$~TeV}, \href{https://arxiv.org/abs/2603.18904}{arXiv:2603.18904} (2026).

\bibitem{promptomega}
A.~S.~Triolo for the ALICE Collaboration, \emph{ALICE charming take on strangeness enhancement in pp collisions}, EPJ Web Conf. \textbf{364} (2026) 11002, \href{https://doi.org/10.1051/epjconf/202636411002}{doi:10.1051/epjconf/202636411002}.

\bibitem{hypertriton}
ALICE Collaboration, \emph{Wave-Function Femtometry: Hypertriton - The Ultimate Halo Nucleus}, \href{https://arxiv.org/abs/2604.07949}{arXiv:2604.07949} (2026).

\bibitem{ppLambda}
ALICE Collaboration, \emph{Unveiling the strong interaction among hadrons at the LHC}, \href{https://doi.org/10.1103/PhysRevLett.131.102302}{Phys. Rev. Lett. 131, 102302} (2023).

\bibitem{ppK}
ALICE Collaboration, \emph{Study of the p--p--K$^+$ and p--p--K$^-$ dynamics using the femtoscopy technique}, \href{https://doi.org/10.1140/epja/s10050-023-01139-9}{Eur. Phys. J. A 59, 298} (2023).

\end{thebibliography}
\end{document}